\documentclass[12pt]{article}
\usepackage[latin1]{inputenc}
\usepackage[T1]{fontenc}
\usepackage{amsmath, amsthm, amsfonts, amssymb, mathrsfs}
\usepackage{epsfig,caption,graphicx}

\begin{document}
\newcommand{\be}{\begin{equation}}
\newcommand{\ee}{\end{equation}}
\newcommand{\ba}{\begin{eqnarray}}
\newcommand{\ea}{\end{eqnarray}}
\newcommand{\nn}{\nonumber}
\newcommand{\bbf}{\mathbf}
\newcommand{\rrm}{\mathrm}
\newcommand{\llp}{\left|+\right\rangle}
\newcommand{\lln}{\left|-\right\rangle}
\newcommand{\rrp}{\left\langle+\right|}
\newcommand{\rrn}{\left\langle-\right|}

\title{Motion of free spins and NMR imaging without a radio-frequency magnetic field.}

\author{$^{a,c}$Kees van Schenk Brill
\footnote{E-mail address: keesvsb@gmail.com},
$^{b}$Jassem Lahfadi
\footnote{E-mail address: lahfadijas@yahoo.fr},
$^{c}$Tarek Khalil
\footnote{E-mail address: tkhalil@ul.edu.lb},\\
$^{a}$Daniel Grucker
\footnote{E-mail address: d.grucker@unistra.fr}\\}
\maketitle
\begin{center}
$^{a}$Laboratoire Icube, UMR 7357, FMTS/CNRS/UdS, Universit\'e de Strasbourg, France\\
$^{b}$ D\'epartement de Physique, Facult\'e des Sciences(V), Universit\'e Libanaise, Nabatieh, Liban\\
$^{c}$Institut de Recherche Math\'ematique Avanc\'ee, UMR 7501 CNRS/UdS,Universit\'e de Strasbourg, France\\

\end{center}

\begin{abstract}
\noindent
NMR imaging without any radio-frequency magnetic field is explained by a quantum treatment of independent spin~$\tfrac 12$. The total magnetization is determined by means of their individual wave function. The theoretical treatment, based on fundamental axioms of quantum mechanics and solving explicitly the Schr\"{o}dinger equation with the kinetic energy part which gives the motion of free spins, is recalled. It explains the phase shift of the spin noise spectrum with its amplitude compared to the conventional NMR spectrum.  Moreover it explains also the relatively good signal to noise ratio of NMR images obtained without a RF pulse.  This derivation should be helpful for new magnetic resonance imaging sequences or for developing quantum computing by NMR.
 
\end{abstract}


\section*{Introduction}
The work of Norbert M\"{u}ller and Alexej Jerschow  \cite{ Muller2006} on nuclear spin noise imaging has reactivated the fundamental idea given by Felix Bloch  \cite{Bloch1946} in 1946 that N nuclei with magnetic moment $\mu$ can be measured because of ``statistically  incomplete cancellation''. A recent review on nuclear spin noise which corresponds to this idea can be found in reference  \cite{ Muller2013}. The aim of this work is to set the frame for a quantum mechanical description of an ensemble of $N$~identical non interacting spins~$\tfrac 12$ at finite temperature which experience the presence of main static and radio frequency (RF) magnetic fields in magnetic resonance experiments. We will recall the elementary quantum description of such a system which explains the small induced current by an ensemble of spins at equilibrium and the possibility to make NMR images without any RF magnetic field only with time dependent gradient magnetic fields. This derivation should be helpful for new magnetic resonance imaging sequences or for developing quantum computing by NMR.
\section{Theoretical description}
\subsection{One spin $\tfrac 12$ in a static magnetic field}
For one nuclear spin~$\tfrac 12$ in a $\vec{B}_0$~magnetic field, the potential part of the Hamiltonian is:
\begin{align}\label{eq1}
\mathscr{H}&=\vec{\mu} \cdot \vec{B}_0
=-\gamma \vec{I}\cdot\vec{B}_0,
\end{align}
where $\gamma$ is the gyromagnetic ratio of the nuclear spin and $\vec{I}$ is the spin that verifies the following equation:
\begin{equation}
\vec{I}=\frac 12\hbar\vec{\sigma}, 
\end{equation}
with $\vec{\sigma}$ a vector defined by the following Pauli matrices:
\begin{align}
\sigma^X=\begin{pmatrix}
            0  & 1  \\
	    1  & 0  \\
           \end{pmatrix},&&
\sigma^Y=\begin{pmatrix}
            0  & -i  \\
	    i  & \phantom{-}0  \\
           \end{pmatrix},&&
\sigma^Z=\begin{pmatrix}
            1  & \phantom{-}0  \\
	    0  & -1  \\
           \end{pmatrix}.
\end{align}
Matrix notation of equation~\eqref{eq1} is given by:
\begin{align}\label{eq2}
\mathscr{H} &= -\tfrac 12 \gamma \hbar 
\begin{pmatrix}
B_Z  & B_X-iB_Y  \\
B_X+iB_Y  & -B_Z  \\
\end{pmatrix}.
\end{align} 
By convention, the vector~$\vec{B}_0=\left(B_X,B_Y,B_Z\right)$ is placed in the $Oz$-direction, that is $\vec{B}_0=\left(0,0,B_0\right)$ and the $xOy$-plane is called the transverse plane. Therefore the matrix form of the Hamiltonian reduces to:
\begin{align} \label{H1}
\mathscr{H} &= -\tfrac 12 \gamma \hbar 
\begin{pmatrix}
B_0  & \phantom{-}0  \\
0  & -B_0  \\
\end{pmatrix}.
\end{align}
The two eigenvalues of the Hamiltonian~$\mathscr{H}$, which give the energy of the quantum states, are: $E_- = -\frac 12 \gamma \hbar B_0$ and $E_+ =  \frac 12 \gamma \hbar B_0$ which have the following two corresponding eigenvectors:
\begin{align}
\left\lvert-\right\rangle= 
\begin{pmatrix}
0 \\
1 \\
\end{pmatrix},&&
\left\lvert+\right\rangle= 
\begin{pmatrix}
1 \\
0 \\
\end{pmatrix}.
\end{align}
The wave function of one spin$\tfrac 12$ at time $ t=0$ is:
\begin{align}
\label{wavefunction}
\bigl\lvert\psi(0)\bigr\rangle& =x_1 \left\lvert-\right\rangle +x_2\left\lvert+\right\rangle= r_1 e^{i\phi_1} \left\lvert-\right\rangle + r_2 e^{i\phi_2}\left\lvert+\right\rangle\\ \nonumber
&=
\begin{pmatrix}
r_2e^{i \phi_2} \\
r_1e^{i \phi_1} \\
\end{pmatrix}
\text{with} \;  r_1^2+r_2^2=1
\end{align}
The 1 indice corresponds to the low energy or ground level.
Time evolution of this quantum system (far from the speed of light) is given by the Schr\"{o}dinger equation: 
\begin{align}
\frac{i\hbar \partial\bigl\lvert\psi(t)\bigr\rangle}{\partial t}&= \mathscr{H}\bigl\lvert\psi(t)\bigr\rangle. 
\end{align}
If we write
\begin{align}
\bigl\lvert\psi(t)\bigr\rangle&=
\begin{pmatrix}
  x_2(t)\\
  x_1(t)\\
\end{pmatrix},
\end{align}
then we have the following differential equations:
\begin{align}\label{Dif1}
\begin{split}
 i\hbar
 \begin{pmatrix}
  \dot{x}_2(t)\\
  \dot{x}_1(t)
 \end{pmatrix}&=-\tfrac 12 \gamma \hbar 
\begin{pmatrix}
B_0  &\phantom{-} 0  \\
0  & -B_0  \\
\end{pmatrix}\cdot
\begin{pmatrix}
x_2(t)\\
x_1(t)\\
\end{pmatrix}\\
&=\tfrac 12\hbar\begin{pmatrix}
   \phantom{-} \omega_0 x_2(t)\\
   -\omega_0 x_1(t)\\
  \end{pmatrix},
\end{split}\\
\intertext{where} 
\omega_0&=-\gamma B_0 
\end{align}
is the resonance or Larmor frequency. The following time dependent wave function is the obvious solution of this system of equations:
\begin{align}
\label{1spinB0}
\bigl\lvert\psi(t)\bigr\rangle &= 
\begin{pmatrix}
r_2 e^{-i\bigl(\tfrac{\omega_0 t}{2}-\phi_2\bigr)}\\
r_1 e^{+i\bigl(\tfrac{\omega_0 t}{2}+\phi_1\bigr)}\\ 
\end{pmatrix} \\ \nonumber
&=
\begin{pmatrix}
 e^{-i\tfrac{\omega_0 t}{2}} & 0\\
 0 & e^{+i\tfrac{\omega_0 t}{2}}\\ 
\end{pmatrix}
\bigl\lvert\psi(0)\bigr\rangle
&=E(\omega_0, t) \bigl\lvert\psi(0)\bigr\rangle
\end{align}
which defines the spin evolution time operator $E(\omega_0, t)$.
\subsection{Effect of a RF magnetic field}
In NMR, a transition between the two states $\left\lvert-\right\rangle$ and $\left\lvert+\right\rangle$ is obtained by a $B_1$~magnetic field rotating in the transverse plane. $B_1$ is assumed to lie along the $Ox$-direction.
\noindent This field is generated by an electromagnetic RF~wave obtained by an oscillating current in a solenoid surrounding the spin sytem. In this case the Hamiltonian has non-diagonal elements due to the RF~magnetic field~$\boldsymbol{\vec{B}_1}$ rotating around~$\boldsymbol{\vec{B}_0}$ with an angular velocity~$\omega$:
\begin{align}
\mathscr{H}_{\mathrm{RF}}(t) &= - \frac 12 \gamma \hbar 
\begin{pmatrix}
B_0  &B_1e^{-i\omega t} \\
B_1e^{i\omega t}  & -B_0  \\ 
\end{pmatrix}.
\end{align} 
In this case time evolution is no longer trivial. We have the following differential equations:
\begin{align}
\begin{split}
\frac{i\hbar \partial\bigl\lvert\psi_{\mathrm{RF}}(t)\bigr\rangle}{\partial t}
&=\mathscr{H}_{\mathrm{RF}}(t)
\begin{pmatrix}
x_2(t) \\
x_1(t) \\ 
\end{pmatrix}\\
&=\tfrac 12\hbar
\begin{pmatrix}
\omega_0  &\omega_1e^{-i\omega t} \\
\omega_1 e^{i\omega t}  & -\omega_0  \\ 
\end{pmatrix}\cdot
\begin{pmatrix}
x_2(t) \\
x_1(t) \\ 
\end{pmatrix},
\end{split}\\
\intertext{where}
\omega_1&=- \gamma B_1.
\end{align}
In the time-independent case ($\omega_1=0$) these equations reduce to equations~\eqref{Dif1}. For the time-dependent case we need to solve the following differential equations:
\begin{align}\label{Dif2}
\begin{pmatrix}
  \dot{x}_2(t)\\
  \dot{x}_1(t)\\
 \end{pmatrix}&=-\frac i2
 \begin{pmatrix}
  \omega_0 x_2(t)+\omega_1e^{-i\omega t} x_1(t)\\
  \omega_1e^{i\omega t} x_2(t)-\omega_0 x_1(t)\\
 \end{pmatrix}.
\end{align}
To solve these two equations we make the following substitutions:
\begin{subequations}
\begin{align}
p(t)&=x_2(t)e^{ i\tfrac{\omega_0}2 t},\\
q(t)&=x_1(t)e^{-i\tfrac{\omega_0}2 t}. 
\end{align}
\end{subequations}
Therefore we have:
\begin{subequations}
\begin{align}
\dot{p}(t)&=\left(-\frac i2\left(\omega_0 x_2(t)+\omega_1e^{-i\omega t} x_1(t)\right)+\frac{i\omega_0}2 x_2(t)\right) e^{i\tfrac{\omega_0}2 t}\\
&=-\frac {i\omega_1}2 e^{-i\omega t} x_1(t) e^{i\tfrac{\omega_0}2 t}=-\frac {i\omega_1}2 q(t) e^{i(\omega_0-\omega) t},\nonumber\\
\intertext{and}
\dot{q}(t)&=\left(-\frac i2\left(\omega_1e^{i\omega t} x_2(t)-\omega_0 x_1(t)\right)-\frac{i\omega_0}2 x_1(t)\right)e^{-i\tfrac{\omega_0}2 t}\\
&=-\frac{i\omega_1}2 e^{i\omega t} x_2(t) e^{-i\tfrac{\omega_0}2 t}=-\frac{i\omega_1}2 p(t) e^{-i(\omega_0-\omega)t}\nonumber.
\end{align}
\end{subequations}
When we take the second derivative of $p(t)$ we obtain the following second order differential equation:
\begin{align}
\begin{split}
\ddot{p}&=-\frac{i\omega_1}2\dot{q}e^{i(\omega_0-\omega) t}+i(\omega_0-\omega)\dot{p}\\
&=-\frac{i\omega_1}2 \left(-\frac{i\omega_1}2 p e^{-i\bigl((\omega_0-\omega)t\bigr)}e^{i(\omega_0-\omega) t}\right)+i(\omega_0-\omega)\dot{p}\\
&=i(\omega_0-\omega)\dot{p}-\frac{\omega_1^2}4 p.
\end{split}
\end{align}
Let $\lambda_{\pm}$ be the solutions of the equation
\begin{equation}
\label{deriv2}
 \lambda^2-i(\omega_0-\omega)\lambda+\frac{\omega_1^2}4=0.
\end{equation}
We have
\begin{align}
 \lambda_{\pm}=i\frac{\Omega \pm \Delta}2 && \text{with} &&\Omega=\omega_0-\omega && \Delta= \sqrt{\Omega^2+\omega_1^2}
\end{align}
and
\begin{align}
p(t)&=C_1 e^{\lambda_{+}t}+C_2 e^{\lambda_{-} t}, 
\end {align}
therefore:
\begin{subequations}
\begin{align}
\begin{split}
x_2(t)&=p(t)e^{-\tfrac{i\omega_0 t}2}\\
&=\left(C_1e^{\lambda_{+}t}+C_2 e^{\lambda_{-} t}\right)e^{-\tfrac{i\omega_0 t}2}\\
&=e^{-i\tfrac{\omega}2t}\left(C_1 e^{i\tfrac{\Delta}2 t}+
  C_2 e^{-i\tfrac{\Delta}2 t}\right),
\end{split}\\
\intertext{with initial value}
x_2(0)&=r_2 e^{i\phi_2}. 
\end{align}
\end{subequations}
A similar computation for $x_1(t)$ leads to:
\begin{subequations}
\begin{align}
  x_1(t)&=e^{\tfrac{i\omega t}2}\left(C_3 e^{i\tfrac{\Delta}2 t}+
  C_4 e^{-i\tfrac{\Delta}2 t}\right),\\
  x_1(0)&=r_1 e^{ i\phi_1}.
\end{align}
\end{subequations}
As $x(t),y(t)$ verify initial values and the differential equations~\eqref{Dif2}  we have:
\begin{align}
 C_1+C_2=r_2 e^{i\phi_2},&&
 C_3+C_4=r_1 e^{ i\phi_1},\\  \nonumber
 (\Delta+\Omega) C_1+\omega_1  C_3=0,&&
 (\Delta-\Omega) C_2-\omega_1  C_4=0
\end{align}
Computation of the constants~$C_i$ is now straightforward and gives: 
\begin{subequations}
\label{Ccoef}
\begin{align}
C_1&= \tfrac 1{2}\left((1-\tfrac \Omega{\Delta})r_2 e^{i\phi_2}-\tfrac{\omega_1}{\Delta}r_1 e^{i\phi_1}\right),\\
C_2&= \tfrac 1{2}\left((1+\tfrac \Omega{\Delta})r_2 e^{i\phi_2}+\tfrac{\omega_1}{\Delta}r_1 e^{i\phi_1}\right),\\
C_3&= \tfrac 1{2}\left(-\tfrac{\omega_1}{\Delta}r_2 e^{i\phi_2}+(1+\tfrac \Omega{\Delta})r_1 e^{i\phi_1}\right),\\
C_4&= \tfrac 1{2}\left(\tfrac{\omega_1}{\Delta }r_2 e^{i\phi_2}+(1-\tfrac \Omega{\Delta})r_1 e^{i\phi_1}\right). 
\end{align}
\end{subequations}
These expressions are simplified if the angular velocity~$\omega$ of the RF~magnetic field verifies the resonance condition $\omega=\omega_0$ and in this case we have $\Delta=\omega_1$, $\Omega=0$.\\
In general case we can write the evolution of the spin wave function in matrix notation as:
\begin{align}
\bigl\lvert\psi_{\mathrm{RF}}(t)\bigr\rangle&=A(\omega,\omega_0,\omega_1,t) \bigl\lvert\psi_{1}(0)\bigr\rangle.
\end{align}
where A matrix is
\begin{align}\label {RFmat1}
A=
\begin{pmatrix}
 e^{-i\tfrac{\omega}2 t} a &  e^{-i\tfrac{\omega}2 t}b\\
 e^{i\tfrac{\omega}2 t}  b &  e^{i\tfrac{\omega}2 t} \bar{a}  
\end{pmatrix}
\intertext{with}
a=\cos \tfrac{\Delta t}2-\tfrac{i\Omega}{\Delta} \sin \tfrac{\Delta t}2
&& b= -i \tfrac{\omega_1 }{\Delta} \sin \tfrac{\Delta t}2
\end{align}
where $\bar{a}$ is the  complex conjugate of $a$.
In this equation it is possible to separate the evolution due to the RF frequency $\omega$. We then obtain:
\begin{align}
\label {RFmat2}
A(\omega,\omega_0,\omega_1,t)&=E(\omega,t)\cdot R(\omega,\omega_0,\omega_1,t),
\end{align}
or
\begin{align}
A&=\begin{pmatrix}
e^{-\tfrac{i\omega t}2}&0\\
0&e^{\tfrac{i\omega t}2}\\ 
\end{pmatrix}
\begin{pmatrix}
 a &  b\\
  b &   \bar{a}  
\end{pmatrix}
\end{align}
At the resonance frequency, the matrix $A$ reduces to:
\begin{align}
\label{Areson}
A_0(\omega_0,\omega_1,t)&=
\begin{pmatrix}
e^{-\tfrac{i\omega_0 t}2}&0\\
0&e^{\tfrac{i\omega_0 t}2} 
\end{pmatrix}
\begin{pmatrix}
\cos\tfrac{\omega_1 t}2&-i \sin\tfrac{\omega_1 t}2\\[0.5em]
-i \sin\tfrac{\omega_1 t}2&\cos\tfrac{\omega_1 t}2 
\end{pmatrix}.
\end{align}
Effect of an RF~pulse is usually described  as a rotation \cite{coh} with an angle of $\theta_1=\omega_1 t$ in the spin space around vector $\boldsymbol{\vec{u}}=\left(u_x,u_y,u_z \right)$ as:
\begin{align}
\boldsymbol{R_{u,\theta_1}^{\tfrac 12}}&=
\begin{pmatrix}
\cos\tfrac{\theta_1}2-iu_z \sin\tfrac{\theta_1}2 & \left( -iu_x-u_y\right) \sin\tfrac{\theta_1}2 \\[0.5em]
\left( -iu_x+u_y\right)\sin\tfrac{\theta_1}2 & \cos\tfrac{\theta_1}2+iu_z \sin\tfrac{\theta_1}2 
\end{pmatrix}.
\end{align}
If  $\boldsymbol{\vec{u}}$ lies in the transverse plane: $\boldsymbol{\vec{u}}=\left( 1, 0, 0 \right)$ where $x$-direction is given by the phase of the RF pulse (in other words by the beginning of the RF pulse) and therefore the rotation matrix becomes:
\begin{align}
\label{eq5}
\boldsymbol{R_{(1,0,0),\theta_1}^{\tfrac 12}}&=
\begin{pmatrix}
\cos\tfrac{\theta_1}2 & -i\sin\tfrac{\theta_1}2 \\[0.5em]
-i\sin\tfrac{\theta_1}2 & \cos\tfrac{\theta_1}2 
\end{pmatrix}.
\end{align}
We notice that this is exactly the transformation matrix that we have computed at resonance in equation  \eqref{Areson}, except for the fact that the time evolution $E(\omega,t)$ is missing in this equation. Therefore in the general case, equation \eqref{RFmat1} gives the evolution of the wave function and equation \eqref{RFmat2} shows that the evolution of the spin system is a time evolution given by matrix $E(\omega,t)$ which depends on the RF frequency followed by a rotation of angle $\theta_1=\Delta t= \sqrt{(\omega_0-\omega)^2+\omega_1^2}\ t$ around vector:
\begin{align}
\boldsymbol{\vec{u}_\omega}: u_x&=\dfrac{\omega_1} {\sqrt{(\omega_0-\omega)^2+\omega_1^2}},\ u_y=0, \\ \nonumber
\ u_z& =\dfrac{\omega_0-\omega} {\sqrt{(\omega_0-\omega)^2+\omega_1^2}}. 
\end{align}
It is important to note that this is a non-commutative product. In other words, the spin evolution in presence of an RF field is, according to equations \eqref{RFmat1} and \eqref{RFmat2}, a matrix product of a time evolution operator and a rotation operator. The wave function after the pulse is given by:             
\begin{align}
\label{1spinRF}
\bigl\lvert\psi_{\mathrm{RF}}(t)\bigr\rangle&=A\cdot\bigl\lvert\psi_{1}(0)\bigr\rangle
=E\cdot R \cdot  \bigl\lvert\psi_{1}(0)\bigr\rangle
\end{align}
which is the same as the equation derived in Abragam textbook \cite{abragam} but here we have calculated explicitly the A matrix. Therefore at resonance the wave function at the end of the RF pulse is equal to the wave function multiplied by the rotation matrix of angle $\theta_1= \omega t$ which is then multiplied by the evolution matrix but there is no time sequence between these two operations. If this elementary description of the spin dynamic retrieve the main characteristic of NMR experiments, it is not complete.
\subsection{Kinetic part of the Hamiltonian}
As an Hamiltonian corresponds to the sum of the potential and the kinetic energies of the physical system under investigation, the potential energy ($-\vec{\mu} \cdot \vec{B}_0$) must be completed by  the kinetic energy ($p^2/2m$)  or more precisely  $ K'=\frac {-\hbar^2}{2m} \mathbf {\nabla^2}(F(r,t)) $ where F is a function of space and time dependent on the spin system. For our purpose we consider only the time dependence $K'(t)$ where $K'$ is a scalar dependent on the absolute temperature for a macroscopic sample. It is well known that such a system obeys always the  Schr\"{o}dinger equation. The complete Hamiltonian for one spin is:
\begin{align}
\mathscr{H}_{T}(t) & =K' -\vec{\mu} \cdot \vec{B}_0\\ \nonumber
&=\tfrac 12\hbar
\begin{pmatrix}
\omega_0 +\frac{2K'}{\hbar} &\frac{2K'}{\hbar} \\
\frac{2K'}{\hbar}  & -\omega_0+\frac{2K'}{\hbar}  \\ 
\end{pmatrix}
\end{align}
and  the Schr\"{o}dinger equation becomes:
\begin{align}\label{Dif2T}
\begin{pmatrix}
  \dot{x}_{2T}(t)\\
  \dot{x}_{1T}(t)\\
 \end{pmatrix}&=-\frac i2
 \begin{pmatrix}
 ( \omega_0+K) x_{2T}(t)+K x_{1T}(t)\\
K x_{2T}(t)-(\omega_0-K)  x_{1T}(t)\\
 \end{pmatrix}.
\end{align}
with $K=2K'/\hbar$.\\
Here to solve these equations we make the following substitutions:
\begin{align}
\label{kinetic1}
p(t)&=x_2(t)e^{i \tfrac{\omega_0 +K}{2}t},
&& q(t)=x_1(t)e^{-i\tfrac{\omega_0 -K}{2}t}. 
\end{align}
which gives  (if $\omega_0$  and $K$ are time  independent):
\begin{align}
\label{kinetic2}
\ddot{p}(t)-i\omega_0\dot{p}(t)+\frac{K^2}4 p(t)=0,
&&\ddot{q}(t)+i\omega_0\dot{q}(t)+\frac{K^2}4q(t)=0
\end{align}
These equations  have the same solutions as equation \eqref{deriv2} with $ \Omega=\omega_0$ and $\omega_1=K$. The solution of these equations is:
\begin{align}
&\begin{pmatrix}
C_1 e^{i\tfrac{\Delta -K}{2}t}+  C_2 e^{-i\tfrac{\Delta+K}{2}t}\\[0.8em]
C_3 e^{i\tfrac{\Delta -K}{2}t}+  C_4 e^{-i\tfrac{\Delta +K}{2}t}
\end{pmatrix}=\\ \nonumber
&\begin{pmatrix}
e^{-i \tfrac{K}{2}t}&0\\
0&e^{-i \tfrac{K}{2}t}\\
\end{pmatrix}
\begin{pmatrix}
C_1 e^{i\tfrac{\Delta}{2}t}+C_2 e^{-i\tfrac{\Delta}{2}t}\\[0.8em]
C_3 e^{i\tfrac{\Delta }{2}t}+C_4 e^{-i\tfrac{\Delta }{2}t} 
\end{pmatrix}
\end{align}
where 
$C_1, C_2,C_3,C_4$ are identical to the coefficients obtained in the previous case and given by equations~\eqref{Ccoef} with exchanging $\Omega$ by $ \omega_0$ and $\Delta$ by $\sqrt{\omega_0^2+K^2}$ .
This solution is again the product of an evolution matrix with a rotation matrix:
\begin{align}
\label{restwave}
\bigl\lvert\psi_{\mathrm{T}}(t)\bigr\rangle&=E_T \cdot R \cdot\bigl\lvert\psi_{1}(0)\bigr\rangle
\end{align}
where $E_T$ is a matrix and not a scalar, and R is given by
\begin{align}
R=\begin{pmatrix}
\cos\tfrac{\Delta t}2-i\tfrac{\omega_0}{\Delta}\sin\tfrac{\Delta t}2&
-i\tfrac{K }{\Delta}\sin\tfrac{\Delta t}2\\[0.5em]
-i\tfrac{K}{\Delta}\sin\tfrac{\Delta t}2&
\cos\tfrac{\Delta t}2+i\tfrac{\omega_0}{\Delta}\sin\tfrac{\Delta t}2
\end{pmatrix}
\end{align}
Therefore the evolution of the wave function of one spin is given by the product of the spin evolution operator and a spin rotation of angle $\theta =\sqrt{\omega_0^2+K^2}t$ around $u(K/\Delta,0, \omega_0/\Delta)$. As the evolution operator is a rotation there is  no oscillation between the two states, and without RF or static magnetic time evolution, it seems that no NMR signal can be detected.

\subsection{General description of one spin  $\tfrac 12$ }
In the general case the Hamiltonian of a spin  $\tfrac 12$ is dependent on the time evolution of the magnetic field $B(t)$ and the kinetic energy $K'(t)$, therefore it can be written as:
\begin{align}
\mathscr{H}_{T}(t) & =-\tfrac 12 \hbar
\begin{pmatrix}
\gamma B_Z +K' &\gamma B_X -i \gamma B_Y +K' \\
\gamma B_X + i \gamma B_Y +K' & -\gamma B_Z +K' \\ 
\end{pmatrix}
\end{align}
and the Schr\"{o}dinger equation becomes:
\begin{align}
\begin{pmatrix}
  \dot{x}_{2}\\
  \dot{x}_{1}\\
 \end{pmatrix}&=-\frac i2
 \begin{pmatrix}
 ( \omega_Z+K) x_{2}+(\omega_X-i\omega_Y+K) x_{1}\\
(\omega_X+i\omega_Y+K) x_{2}-(\omega_Z-K)  x_{1}\\
 \end{pmatrix}
\end{align}
with $K=2 K'/ \hbar$,  $ \omega_X=-\gamma B_X$,  $\omega_Y=-\gamma B_Y$ ,  $\omega_Z=-\gamma B_Z $ .\\
We look for an adequate exponential substitution $p(t)=x_2(t)e^{A(t) }$ and  $q(t)=x_1(t)e^{B(t)}$ which derivation gives:
\begin{subequations}
\label{general1}
\begin{align}
\dot{p}&= p \left[ \dot{A} - \tfrac i2(\omega_Z+K) \right] - \tfrac i2 ( \omega_X -i\omega_Y+K) q e^{(A-B)}\\
\dot{q}&= q \left[ \dot{B} + \tfrac i2(\omega_Z-K) \right] - \tfrac i2 ( \omega_X +i\omega_Y+K) p e^{(B-A)}
\end{align}
\end{subequations}
For $A=\tfrac i2(\omega_Z(0)+K(0))t$, $B=-\tfrac i2(\omega_Z(0)-K(0))t$ these equations simplify but the second derivation introduce a coupling between the two states:
\begin{subequations}
\label{general2}
\begin{align}
\ddot{p}=&-\tfrac {1}4 ( \omega_X^2+\omega_Y^2+K^2+2K\omega_X)p + \dot{p} (\dot{A}-\dot{B)} \\ \nonumber
&- \tfrac{i}2  (\dot{\omega}_X-i\dot{\omega}_Y+\dot{K}) q e^{(A-B)}\\ 
\ddot{q}=&-\tfrac {1}4 ( \omega_X^2+\omega_Y^2+K^2+2K\omega_X)q + \dot{p} (\dot{B}-\dot{A)}  \\ \nonumber
&- \tfrac{i}2  (\dot{\omega}_X-i\dot{\omega}_Y+\dot{K}) p e^{(B-A)}
\end{align}
\end{subequations}
The constant solution of these equations corresponds to:
\begin{subequations}
\begin{align}
\label{general3}
\ddot{p}&-i \omega_Z \dot{p}+\tfrac 14(\omega_X^2+\omega_Y^2+K^2+2K\omega_X)p=0  \\ 
\ddot{q}&+i \omega_Z \dot{q}+\tfrac 14(\omega_X^2+\omega_Y^2+K^2+2K\omega_X)q=0 
\end{align}
\end{subequations}
In case of constant magnetic field $B(0,0,B_0)$ we retrieve equations \eqref{kinetic2} and the comparison with these equations shows that the general derivation introduces a coupling of the kinetic energy and the X component of the magnetic field in case of non constant magnetic fields. The solutions of equations \eqref{general3} is, as usually, obtained by the constant solution then varying the constants.\\
The constant solution of these equations, equivalent to equation \eqref{deriv2}, is :
\begin{subequations}
\label{general4}
\begin{align}
x_2(t)&=p(t)e^{-A(t)} =\left(C_1 e^{\lambda_{+}t}+C_2 e^{\lambda_{-} t}\right)e^{-\tfrac i2(\omega_Z+K)t}\\
x_1(t)&=q(t)e^{-B(t)} =\left(C_3 e^{-\lambda_{-}t}+C_4 e^{-\lambda_{+} t}\right)e^{\tfrac i2(\omega_Z-K)t}
\end{align}
\end{subequations}
with
\begin{align}
\label{general5}
 \lambda_{\pm}&=\frac{i\left(\omega_z\pm \Delta\right)}2, \\ \nonumber
  \Delta&= \sqrt{\omega_z^2+\omega_1^2} ,
  && \omega_1^2= \omega_X^2+\omega_Y^2+K^2+2K\omega_X
\end{align}
where $C_i$ coefficients are given by equations \eqref{Ccoef} with $\Omega=\omega_0-\omega=\omega_z$. It is important to notice that $\omega_z$ is now the frequency difference with the reference frequency $\omega_0$ which is the resonant frequency at time $t=0$. The explicit constant solution at each time $t$ is:
\begin{subequations}
\begin{align}
\label{general6}
x_2(t)&=r_2\left(  \cos \tfrac{\Delta}2 t \cos(-\tfrac{K}2 t+ \phi_2)-\tfrac{\omega_z}{\Delta } \sin \tfrac{\Delta}2 t \sin(\tfrac{K}2 t- \phi_2)\right)\\
\nonumber
&+r_1 \tfrac{\omega_1}{\Delta } \sin \tfrac{\Delta}2 t  \sin (-\tfrac{K}2 t +\phi_1)\\ \nonumber
 &-i\left( r_1 \tfrac{\omega_1}{\Delta } \sin \tfrac{\Delta}2 t \cos(-\tfrac{K}2 t+ \phi_1)  +  r_2( \tfrac{\omega_z}{\Delta} \sin \tfrac{\Delta}2 t \cos(\tfrac{K}2 t- \phi_2)\right)\\
 \nonumber
 &-i\left( \cos \tfrac{\Delta}2 t \sin(-\tfrac{K}2 t+ \phi_2) \right) \\
 \label{general6b}
  x_1(t)&=r_1\left(  \cos \tfrac{\Delta}2 t \cos(-\tfrac{K}2 t+ \phi_1)-\tfrac{\omega_z}{\Delta } \sin \tfrac{\Delta}2 t \sin(-\tfrac{K}2 t+ \phi_1)\right)\\
  \nonumber
  &-r_2 \tfrac{\omega_1}{\Delta } \sin \tfrac{\Delta}2 t  \sin (\tfrac{K}2 t -\phi_2)\\ \nonumber
  &+ i \left( r_1( \tfrac{\omega_z}{\Delta} \sin\tfrac{\Delta}2 t \cos(-\tfrac{K}2 t+ \phi_1)+  \cos \tfrac{\Delta}2 t \sin(-\tfrac{K}2 t+ \phi_1) )\right)\\
  \nonumber
  & -  i \left( r_2 \tfrac{\omega_1}{\Delta } \sin \tfrac{\Delta}2 t \cos(\tfrac{K}2 t- \phi_2)  \right)
 \end{align}
 \end{subequations}
which verify $\vert x_1(0)\vert ^2=r_1^2$, $\vert x_2(0)\vert ^2=r_2^2$ and $\vert x_1(t)\vert ^2 + \vert x_2(t)\vert ^2=1$. 
\subsection{Two spins $\tfrac 12$}
For two spins we consider a four dimensional Hilbert space which is a tensor product of the two dimensional Hilbert spaces of the one spin system. As the energy is given by the eigenvalues of the Hamiltonian, energy conservation needs that the eigenvalues of the sum matrix is equal to the sum of the  eigenvalues of each Hamiltonian matrix. This property is obtained by the Kronecker sum $\oplus_K$ of two matrices which is defined by:
\begin{equation}
A \oplus_K B = A\otimes I_n + I_m \otimes B,
\end{equation} 
with $\otimes$ the tensor product with identity matrices $I$ of dimensions $n$ the first dimension of matrix $B$ and $m$ the second dimension of matrix $A$.\\
Therefore the Hamiltonian of the two spin system is:
\begin{align}
\label{HamiltonianHomo2}
\mathscr{H}_{2}=\left(K'_1-\boldsymbol{\vec{\mu}_1} \cdot \boldsymbol{\vec{B_1}} \right) \oplus_K \left(K'_2-\boldsymbol{\vec{\mu}_2} \cdot \boldsymbol{\vec{B_2}} \right)
\end{align} 
This description is the most general possible for two independent spins in a $B$ field. The description we have obtained for the wave function of two spins is easily generalized and the Hamiltonian of $N$ spins is just the Kronecker sum of the one spin Hamiltonian:
\begin{align} \label{HamN}
\mathscr{H}_{N}=\bigoplus_{i=1}^N  \mathscr{H}_1.
\end{align}
This property shows the importance of the Hamiltonian of one spin which determines the dynamic of the whole free spin system.
\section{Measurements}
In NMR, the measurement of the spin system is obtained by the current induced in a tuned $LCR$ circuit which corresponds to the photon emission when a spin switch from the excited state to ground state. It is well known that after a $RF$ pulse an electrical current is induced in the $LCR$ circuit, but even in absence of $RF$ pulse, the presence of the sample can be detected by the nuclear-spin noise \cite{Bloch1946}, \cite{Hahn85}, \cite{Hahn87}, \cite{Ernst1989}. This signal has even been used for imaging sample without any RF pulse \cite{Muller2006} but in this case we will see that the presence of magnetic field gradients explains an increase of the NMR signal which permits to obtain images with a relatively good signal to noise ratio.\\
In order to discuss the difference between these two cases, we calculate the probability of a spin to be in each state from equations \eqref{general6} and\eqref{general6b} :
\begin{align}
\label{general7}
x_2 \bar{x_2}&=r_2^2 +\sin \tfrac{\Delta}2 t \left[ r_2^2 A+ r_1^2 \tfrac{\omega_1^2}{\Delta^2 } \sin \tfrac{\Delta}2 t + 2 \tfrac{\omega_1}{\Delta} r_1 r_2 B  \right]\\ \nonumber
x_1 \bar{x_1}&=r_1^2 +\sin \tfrac{\Delta}2 t \left[ r_1^2 A+ r_2^2 \tfrac{\omega_1^2}{\Delta^2 } \sin \tfrac{\Delta}2 t - 2 \tfrac{\omega_1}{\Delta} r_1 r_2 B  \right]\\ \nonumber
 &\text{with}\\ \nonumber
 A&=\sin \tfrac{\Delta}2 t ( \tfrac{\omega_z^2}{\Delta^2} - 1)\\  \nonumber
 B&=(\cos \tfrac{\Delta}2 t \sin(\phi_1- \phi_2) + \tfrac{\omega_z}{\Delta}\sin \tfrac{\Delta}2 t \cos(\phi_1- \phi_2)
 \end{align}
which shows that the probability of a spin to be in one state is periodic around the value at t=0 and the period is $\tfrac {4 \pi}{\Delta}$. Therefore the probability for a spin to switch from a state to the other  depends on the value of $\Delta=\sqrt{\omega_z^2+\omega_X^2+\omega_Y^2+K^2+K \omega_X}$. For a free spin in a very homogeneous magnetic field $\Delta=K$ (remember that $\omega_z= \omega_0-\omega(0)=0)$ and the probability for a spin to change from one state to the other at a rate greater than the spontaneous oscillation between the two states is $\tfrac{\Delta}{\omega_0}$. In presence of a RF magnetic field, we are in the case where $\omega_z= \omega_0$ and at resonance the probability to change from one state to the other is 1 because $\omega_X$, $\omega_Y$ and $K$ are small compared to $\omega_0$.

\subsection{Nuclear spin noise experiments}
After the first observation of random noise emission from nuclear spins by  Sleator, Hahn, Hilbert and Clarke \cite{Hahn85}, \cite{Hahn87} using a superconducting SQUID, McCoy and Ernst \cite{Ernst1989} have observed nuclear spin noise at room temperature using a conventional (300 MHz) NMR spectrometer. They compared the proton resonance of the methyl group in ethanol without any RF pulse just by co-adding 8000 spectra with a conventional one pulse spectra. The most  surprising features of the spectrum without any RF is the negative deviation compared to the spectrum obtained by a $\pi/2$ pulse excitation and the amplitude of the signal which is about  $10^8$ time smaller. Our elementary derivation retrieves this main characteristics.\\
Without RF and no field inhomogeneity, the  probability for a spin to emit a photon is given by $-K/\omega_0$, as $K=\tfrac{2K'}{\hbar}$ where $K'$ is the kinetic energy of a proton $K'=\tfrac{\hbar^2}{2m}$ therefore the probability of a NMR signal detection is $-\tfrac{\hbar}{m \omega_0}= 2 \, 10^{-16}$ for protons in a 300 MHz magnetic field as in the McCoy and Ernst experiment. This probability is very small therefore in case of a great number of spins the statistical fluctuations around the mean values of $r_1$ and $r_2$ gives a squared probability which is $\sqrt{2 \, 10^{-16}}\simeq10^{-8}$. The minus sign gives the deviation from the probability to find the spin in a state at $t=0$. In case of a RF magnetic field at resonance the probability is one and the deviation from state  at t=0 is positive. The elementary description of the spin motion retrieves the main characteristics of a spin noise spectrum compared to the spectrum obtained with a $\pi/2$ pulse.
\subsection{NMR imaging experiments}

\begin{figure*}
\includegraphics[scale=0.56]{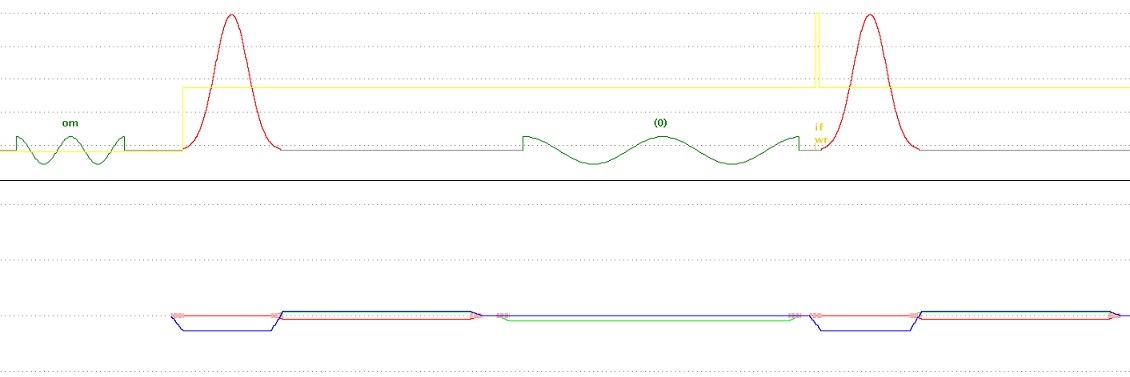}
\caption{\label{fig:1}Conventional gradient echo imaging sequence.}
\end{figure*}

\begin{figure*}
\includegraphics[scale=0.6]{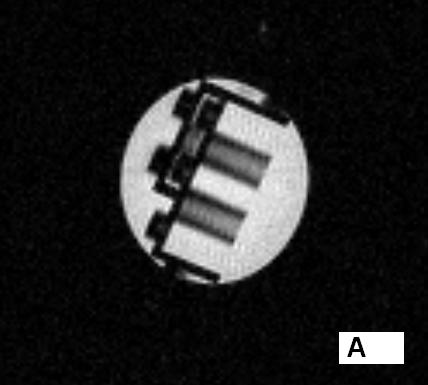}
\includegraphics[scale=0.59]{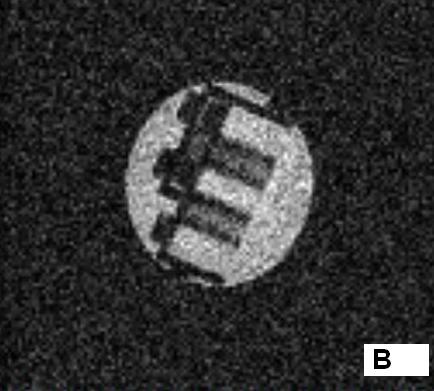}
\caption{\label{fig:2} NMR images of a test object (Lego pieces): A with RF and B without RF pulses.}
\end{figure*}

In imaging experiments, magnetic gradient fields are applied in all space directions as in the work of  M\"{u}ller and Jerschow  \cite{ Muller2006} or in the three main direction as in phase encoding imaging thechnics. In this case a first gradient is applied in the X-direction with a RF pulse in order to select a slice with its thickness, after that a gradient is applied in a orthogonal direction in order give a phase coherence and after that a gradient is applied in the third orthogonal direction during signal acquisition. Figure~\ref{fig:1} gives the schematic of one of the simplest imaging sequence. 
The first line corresponds to the RF channel where ``om'' is the synchronization of the lock-in detector with RF, the gaussien pulse shape is the red line and (o) corresponds to the acquisition of the current induced in the NMR receiver. The second line corresponds to the applied magnetic field gradients in blue for the $Z$ (or slice) direction, in red for the $Y$ (or phase) direction and in green for the $X$( or read) direction. When applying gradients the exact synchronization on the reference frequency $\omega_0$ is mandatory to have the $t=0$ values and the same X-direction for each phase encoding step or gradient orientation. Image A of figure ~\ref{fig:2} 
 is the conventional NMR image of a test object composed of two Lego pieces in water with 1g/L $CuSO_4$ and Agar/Agar 10 g/L. Images  of 128x128 pixels was obtained on a Bruker 7T BS70/30 small animal imaging system. Image A of figure (2) was obtained with field of view of 58 mm corresponding to a magnetic field gradient of 2 $10^{-2}$ T/m, a slice thickness of 1.16 mm, and an echo-time of 3.9 ms with a repetition-time of 8 ms corresponding to an acquisition-time of 1s 23ms. The RF pulse used was a Gaussian pulse of 15 mW power corresponding to a flip angle of $5^o$. The signal to noise ratio in image A is 18.1. Image B of figure~\ref{fig:2}  was acquired with the same parameters except that there was no RF pulse (Power excitation pulse <$10^{-9} W$), in this case the signal to noise ratio is 5.0. We see from equations \eqref{general7}  that NMR images can be obtained  without RF magnetic field due to the fact that the probability to emit a photon by the spin system is not very different with or without RF because $\Delta$ without RF is  $\Delta= \sqrt{\omega_Z^2+\omega_X^2 +\omega_Y^2+K^2 +2K\omega_X}$ and with RF  $\Delta= \sqrt{\omega_Z^2+\omega_X^2+\omega_Y^2+K^2+2K(\omega_X+ \omega_{RF} cos\omega t) +\omega_{RF}^2}$.  In effect in presence of magnetic field gradients $\Delta$ is not very different in both cases because $ \omega_{RF}$ is always small compared to $\omega_X,\omega_Y$ and the variation of $ \omega_Z$. An adequate discussion of the enhancement of the NMR signal with and without a RF pulse needs a precise description of the magnetic properties of the system under investigation and the information on spatial distribution of the spins, but the important result is  that NMR imaging with good signal to noise ratio can be obtained without any RF magnetic field. Such imaging sequence can be of great medical interest for decreasing RF power deposition in a patient.

\section*{Conclusion}
This work has shown that an elementary approach of the quantum description of an ensemble of independent spin $\tfrac12$ can retrieve the magnetic resonance basic experimental facts. This derivation completes the density matrix formalism which is well adapted for describing all NMR experiments where spin coupling has important effects. Experiments using nuclear spin noise are difficult to be explained by the matrix density formalism without introducing an extra probability factor and a stochastic operator as in the paper of Field and Bain \cite{field}. A precise description of NMR quantum effects should participate to the quantum mechanic fundamental aspects used in the development of quantum computing where NMR is still the only example of an experimental quantum computer working with more than ten qubits. The main result of our approach is to predict that the quantum interference effects are best detected in very homogeneous magnetic field without any RF pulse. 

\section*{Acknowledgments}
One of us (T.K.) would like to thank the Laboratoire de Physique Th\'eorique of the Universit\'e Louis Pasteur (UMR 7085 ULP/CNRS) for its kind hospitality during the time the present work was completed. D.G. would like to thank Jules Grucker and Jacques Baudon for helpful discussions at the beginning of this work.\\

\end{document}